\documentclass{article}
\usepackage{frascatiphys}

\def\lqcd{\Lambda_{\rm QCD}}
\def\OMIT#1{{}}

\def\vereq#1#2{\lower3pt\vbox{\baselineskip1pt\lineskip1pt
     \ialign{\\$#1\hfill##\hfil\\$\crcr#2\crcr\sim\crcr}}}

\begin{document}

\title{INCLUSIVE DETERMINATIONS OF $|V_{ub}|$ AND $|V_{cb}|$ }

\author{Christian W.~Bauer
\em California Institute of Technology}
\maketitle
\baselineskip=11.6pt

\begin{abstract}
In this talk I review the status of our ability to extract the CKM matrix elements $|V_{ub}|$ and $|V_{cb}|$ from inclusive semileptonic decays. I focus on model independent determinations of these
parameters and discuss the expected theoretical uncertainties. 
\end{abstract}

\baselineskip=14pt

\section{Introduction}
The magnitudes of the Cabbibo, Kobayashi, Maskawa (CKM) matrix elements 
$V_{ub}$ and $V_{cb}$ are two of the parameters of the standard model which 
can be determined at current experimental facilities producing $B$ mesons. 
Semileptonic decays of $B$ mesons mediated by the weak decay of a $b$ quark to either an 
up or a charm quark are an ideal way to perform these measurements, since the part of the 
process involving the leptonic final states can be calculated perturpatively. 
The theoretical calculations required can be split into two parts. First, the 
decay rate of the $b$ quark to either an up or a charm quark is required, and second the 
hadronic effects which bind 
these quarks into the observed hadrons in the experiment have to be dealt with. 

The perturbative expressions for the 
$b \to u \ell \bar \nu$ decay is known to order $\alpha_s^2$\cite{ratebuloop},
while the $b \to c \ell \bar \nu$ decay rate is currently known to order $\alpha_s^2 \beta_0$\cite{ratebcloop}, where $\beta_0$ is the one loop coefficient of the QCD 
beta function. The hadronization effects can not be calculated perturbatively and is governed
by long distance physics. There are two 
distinct ways to extract the CKM from decays of $B$ mesons. One can use exclusive decays
to a well defined hadronic final state, such as $D$ or $D^*$ mesons for $b \to c$ transitions, or
$\pi$ or $\rho$ mesons for the measurement of $|V_{ub}|$. All non-perturbative physics is then 
encoded in the hadronic form factors. For $D$ and $D^*$ mesons heavy quark effective theory
(HQET)\cite{mwbook} can be used to obtain the form factor at leading order in an expansion in $1/m_{b,c}$ at the zero recoil point\cite{IW}, and because of Luke's theorem\cite{lukestheorem} 
corrections are absent at order $1/m_{b,c}$. For the decay to an up quark HQET is not 
applicable, and the relevant form factors have to be determined using other 
non-perturbative methods, such as lattice QCD\cite{latticereview} or QCD sumrules\cite{sumrulereview}. Recently there has also been progress 
using the soft collinear effective theory\cite{scet} to determine the required form factors from 
experiment.

An alternative approach is to use decays to inclusive final states, which include all final states 
containing either an up or a charm quark. Decays to such inclusive final states can be calculated
using the operator product expansion (OPE), which states that at leading order in $1/m_b$ the
inclusive decay is identical to the perturbatively calculable parton level decay. Corrections are given 
by matrix elements of local operators, which are suppressed by powers in $1/m_b$. By 
determining enough of these matrix elements the CKM parameters $V_{ub}$ and $V_{cb}$ can
be determined with high accuracy. I review the recent progress on inclusive determinations of $V_{ub}$ and $V_{cb}$ in this talk.

\section{Inclusive determination of $V_{\rm \lowercase{ub}}$}


%
The inclusive decay rate $B \to X_u \ell \bar \nu$ is directly proportional to $|V_{ub}|^2$ and can be calculated reliably and with small uncertainties using the operator product expansion (OPE).
Unfortunately, the $\sim$100 times background from $B \to X_c \ell \bar \nu$ makes the 
measurement of the totally inclusive rate an almost impossible task. Several cuts have been 
proposed in order to reject the $b \to c$ background, however care has to be taken to 
ensure that the decay rate in the restricted region of phase space can still be predicted 
reliably theoretically. The proposed cuts are
\begin{enumerate}
\item Cut on the lepton energy $E_\ell > (m_B^2-m_D^2)/(2m_B)$
\item Cut on the hadronic invariant mass $m_X < m_D$\cite{mXcut}
\item Cut on the leptonic invariant mass $q^2 > (m_B-m_D)^2$\cite{qsqcut}
\item Cut on light cone component of the hadronic momentum $P_+ < m_D^2/m_B$\cite{Ppcut}
\item Combined lepton-hadron invariant mass cut\cite{combinedcut}
\end{enumerate}

While the cut on the energy of the charged lepton is easiest to implement experimentally, it
has the largest theoretical uncertainties. This is due to the fact that only $\sim 10\%$ of the
$b \to u$ events survive this cut, amplifying any higher order, uncalculated terms drastically.
Thus, it is not useful for a precision determination of $|V_{ub}|$, although it can be 
used as a check for consistency. 

\begin{figure}[t]
 \vspace{3cm}
\includegraphics{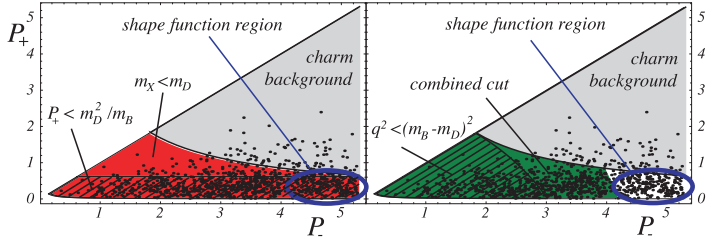}
 \caption{\it
     The dalitz plot in the  $q^2/s_H$ and $q^2/E_\ell$ plane. In both plots the gray area denotes the area contaminated by $b \to c$ events. The left plot shows the $m_X<m_D$ and $P_+ < M_D^2/m_B$ cuts, while the right hand plot shows the $q^2 > (m_B-m_D)^2$ and the combined $q^2-m_X$ cut. Also shown in both plots is the shape function region.
     \label{fig:dalitz}}
\end{figure}

 The remaining four cuts each have their advantages and disadvantages, and it remains to 
be seen which will yield the individually smallest uncertainty on  $|V_{ub}|$ ultimately. To 
illustrate the effect of these four phase space cuts, we show the allowed phase space of
the $B \to X_u \ell \bar \nu$ transition, in terms of two light cone projections of the hadronic four-momentum, 
\begin{eqnarray}
P_+ &=& n \cdot P = E - |\vec{P}|\nonumber\\
P_- &=& n \cdot P = E + |\vec{P}|\,.
\end{eqnarray}
The projections satisfy $P_+ P_- = P^2$ and thus it is obvious that the boundaries
of phase space are 
\begin{eqnarray}
m_\pi^2 / P_- < P_+ < P_- < m_B
\end{eqnarray}
The resulting phase space diagram is shown in Fig.~\ref{fig:dalitz}. Also 
displayed in a rough distribution of the events obtained
from a toy Monte Carlo simulation. While this distribution should not be viewed as a sound
theoretical prediction, it qualitatively helps to understand the phase space better. The region of phase space occupied by the $b \to c$ background is given by $P_+ P_- > m_D^2$ and is indicated by 
the gray area. 

The region satisfying $P_+ \ll P_-$, denoted by the ellipse in Fig.~\ref{fig:dalitz}, is called the 
shape function region. The decay rate in the presence of cuts which 
include this region contain higher dimensional operators contributing at 
order $\left(P_+ \lqcd/P_-^2\right)^n$. This fraction becomes order unity and all these terms
have to be resummed to all orders into an unknown function, called the shape function~\cite{shape}. 
This function is a universal property of the $B$ meson, and can be measured
in other $B$ decays, such as the radiative decay $B \to X_s \gamma$. Note that it is not
simply related to the $b$ quark mass and the kinetic energy of the $b$ quark as is 
often assumed~\cite{shapebm}. In fact, at , leading order 
in both $\alpha_s$ and $\lqcd/m_b$, the shape of the photon energy spectrum is precisely
given by this light cone distribution function. 
At order $1/m_b$ several
new subleading shape functions enter\cite{subleading}, which are at present completely 
unknown. Thus, even with perfect knowledge of the photon energy spectrum in 
$B \to X_s \gamma$ the uncertainties in regions
of phase space which include the shape function region of order $\lqcd/m_b$. 

The regions of phase space surviving the four cuts are also illustrated in Fig.~\ref{fig:dalitz}. On 
the left we show the $m_X < m_D$ and $P_+ < m_D^2/m_B$ cuts, which both include the 
shape function region, while on the right we show the $q^2 > (m_B-m_D)^2$ and the combined 
hadron-lepton invariant mass cut, which do not include the shape function region. It is clear 
that the cut on the hadronic invariant 
mass $m_X < m_D$\cite{mXcut} is optimal in the sense that it keeps all events which are 
not accessible by
$b \to c \ell \bar\nu$ transitions. It has been estimated that $\sim 80\%$ of the $b \to u$ events survive this cut. Uncertainties from subleading shape functions are of order $\lqcd/m_b$, however they have recently been estimated to be at the few
percent level\cite{mxsubleading}. Precise knowledge of the shape function is however still
required to achieve an uncertainty on $|V_{ub}|$ below the 10\% level. 

The situation is similar for the cut on the light cone momentum $P_+$, which also includes the
shape function region. While this cut includes slightly less phase space, it has been argued that
the relationship between the shape function and the differential rate of $B \to X_s \gamma$ is 
slightly simpler for this cut than for the $m_X$ cut described above~\cite{Ppcut}. The resulting uncertainties 
on $|V_{ub}|$ are expected to be at the same order as for the $m_X$ cut. 

The situation is qualitatively different for the remaining two cuts, which involve a cut on 
the leptonic invariant mass. Since a lepton invariant mass cut removes the shape function
region, the decay rate in the presence of these cuts can be calculated using the standard
OPE in an expansion in local operators, but the expansion is in powers of $1/m_c$ rather
than $1/m_b$\cite{neubertqsq}.  For the pure $q^2$ cut, where $q^2 > (m_B-m_D)^2$,
the fraction of events surviving the cut is estimated to be about $(17\pm 3)\%$\cite{qsqcut}. 
This gives an uncertainty on $|V_{ub}|$ at the 10\% level. 

The final cut discussed here is a combined cut on both the hadronic and the leptonic invariant
mass. The idea here is to use the cut on $m_X$ to remove the charm background, and the 
cut on $q^2$ to keep the sensitivity on the shape small. The ideal combination of cuts
remains to be determined in a detailed experimental study, but using the combined cuts
$m_X < m_D \,{\rm GeV}, \, q^2 > 6 \, {\rm GeV}^2$ one finds the fraction of surviving events
to be $(45 \pm 5)\%$\cite{combinedcut}. Since the decay rate is proportional to $|V_{ub}|^2$, this allows for a determination of $|V_{ub}|$ with uncertainties well below the 10\% level. 

To summarize, there are currently five types of cuts to eliminate the charm background
proposed in the literature. While a cut on the lepton energy is easiest to measure, it has by 
far the largest theoretical problems. A cut on the leptonic invariant mass alone also leads
to relatively large theoretical uncertainties and will probably not yield a measurement
of $|V_{ub}|$ with uncertainties below the 10\% level. The remaining three cuts all can 
yield a determination of this CKM matrix element with uncertainties considerably 
below the 10\% level, and 
all of them should be used together for a precision measurement of $|V_{ub}|$.

\section{Inclusive determination of $V_{\rm \lowercase{cb}}$}

Inclusive semileptonic $B$ decays can be calculated using an operator product expansion (OPE). This leads to a simultaneous expansion in powers of the strong coupling constant $\alpha_s(m_b)$  and inverse powers of the heavy $b$ quark mass. At leading order in this expansion this reproduces the parton model result
\begin{eqnarray}
\Gamma_0 = \frac{G_F^2 |V_{cb}|^2 m_b^5}{192 \pi^3} \left( 1-8 \rho + 8 \rho^3 - \rho^4 -12 \rho^2 \log \rho \right)\,,
\end{eqnarray}
where $\rho = m_c^2/m_b^2$, and nonperturbative corrections are suppressed by at least two powers of $m_b$. The state of the art is to use theoretical predictions to order $\alpha_s^2\beta_0$\cite{ratebcloop} in the perturbative expansion, to order $\lqcd^3/m_b^3$\cite{totalm^3} in the non-perturbative power expansion and to 
order $\alpha_s \lqcd/m_b$ in the mixed terms. Here $\beta_0$ is the one loop coefficient of the QCD beta function $\beta_0 = 25/3$ for $n_f=4$ light quark flavors. There are no non-perturbative 
contributions at order $1/m_b$ and thus the inclusive rate can be written schematically as
\begin{eqnarray}\label{Gammaincl}
\Gamma^{b \to c} &=& \Gamma_0 \Bigg\{ 1+ A \left[\frac{\alpha_s}{\pi}\right] + B \left[\left(\frac{\alpha_s}{\pi}\right)^2 \beta_0\right] + 0 \left[\frac{\Lambda}{m_b}\right] + C \left[\frac{\Lambda^2}{m_b^2}\right] \nonumber\\
&& + D \left[\frac{\Lambda^3}{m_b^3}\right] + E \left[\frac{\alpha_s}{\pi}\frac{\Lambda}{m_b} \right]  +{\cal O} \left(\alpha_s^2, \frac{\Lambda^4}{m_b^4}, \alpha_s\frac{\Lambda^2}{m_b^2} \right) \Bigg\}\,,
\end{eqnarray}
The coefficients $A - E$ depend on the quark masses $m_{(c,b)}$. At order $\lqcd^2/m_b^2$ there are
two matrix elements $(\lambda_{1,2})$ parametrizing the non-perturbative physics, while 
at order $\lqcd^3/m_b^3$ there are
six additional matrix elements $(\rho_{1,2}, {\cal T}_{1-4})$. 

The total inclusive branching fraction for $B$ decays is currently measured with uncertainties
around 2\%. To predict this branching ratio with comparable precision requires detailed 
knowledge of the 
value of the matrix elements $\lambda_{1,2}$ and even some rough knowledge of the 
matrix elements at order $\lqcd^3/m_b^3$. The best way to determine these parameters is to
use the semileptonic data itself. Many differential decay spectra have 
been measured, and 
moments of these spectra have been calculated to the same accuracy as the total branching 
ratio itself\cite{moments}. A global fit to all experimental data is able to test how well the OPE is able
to describe the inclusive observables\cite{globalfit1}. 

The mass of the $b$-quark which naturally appears in the OPE calculations is the pole
masse.  It has been long known that using these pole masses gives rise
to a poorly behaved perturbative expansion, due to the presence of a renormalon. There 
are several threshold mass definitions, which do not contain a renormalon, called 
$1S$ mass\cite{1Smass}
PS mass\cite{PSmass} , and kinetic mass\cite{kinmass}. 

The $c$
quark can be treated as a heavy quark. This allows one to compute the $D^{(*)}$ meson masses as
an expansion in powers of $\lqcd/m_c$. The observed $B-D$ mass splitting can be used
to determine $m_b-m_c$. Since the computations are peformed to $\lqcd^3/m_c^3$, this
introduces errors of fractional order $\lqcd^4/m_c^4$ in $m_c$, which gives
fractional errors of order $\lqcd^4/(m_b^2 m_c^2)$ in the inclusive $B$ decay
rates, since charm mass effects first enter at order $m_c^2/m_b^2$. This is the
procedure used in Ref.~\cite{globalfit1}. An alternative approach is to avoid using the 
$1/m_c$ expansion for the charm
quark~\cite{GU}. In this
case heavy quark effective theory (HQET) can no longer be used for the $c$ quark
system, and there are no constraints on $m_c$ from the $D$ and $D^*$ meson
masses. At the same time, it is not necessary to expand heavy meson states in an
expansion in $1/m_{b,c}$, so that the time-ordered products ${\cal T}_{1-4}$ can be
dropped. The number of parameters is the same whether or not one expands in
$1/m_c$.

Currently, there are 75 pieces of data available combining moments of the hadronic invariant
mass spectrum and the lepton energy spectrum of inclusive measured of semileptonic decays and  
he photon energy spectrum in 
$B \to X_s \gamma$ by BABAR\cite{Babardata}, BELLE\cite{Belledata}, 
CDF\cite{CDFdata}, CLEO\cite{Cleodata} and DELPHI\cite{Delphidata}
 together with moments of the photon energy spectrum in 
$B \to X_s \gamma$ measured by BABAR, BELLE and CLEO. These observables can all 
be predicted using the same OPE and have been calculated in all of the mass schemes 
discussed above and depend on 7 parameters. A global fit to all these 75 observables 
was performed in~\cite{globalfit2}. 
This allowed to extract the value of $|V_{cb}|$ 
simultaneously with the non-perturbative parameters of the OPE. It was shown that all schemes give consistent values for $|V_{cb}|$, $m_b$ and the matrix elements appearing at order $1/m_b^2$. 
In Table~\ref{tab:central} we show the results of the fits in the $1S$ and the kinetic scheme. 
One can see that the two schemes give consistent results, with the uncertainties in the $1S$ scheme
being slightly smaller than the ones in the kinetic scheme. 
\begin{table*}[tbp]
\begin{tabular}{c||c|c|c}
~Scheme~  &  
  ~$|V_{cb}|\times 10^3$~  &  ~$m_b^{1S}\, [{\rm GeV}]$~  &  ~~~$\lambda_1\, [{\rm GeV}^2]$~~~  
\\
\hline\hline
$1S_{\rm exp}$ &  
$42.1\pm 0.6$ &
$4.68\pm 0.04$ & 
$-0.23\pm 0.06$  
\\ \hline
kin$_{\rm exp}$&   
$42.2 \pm 0.4 \pm 0.4 $ &
$4.67\pm 0.04 \pm 0.02$ & 
$-0.17\pm 0.06 \pm 0.06$ 
\\ \hline\hline
\end{tabular}
\caption{Fit results for $|V_{cb}|$, $m_b$ and $\lambda_1$  in the
$1S$  and kin  schemes, where $m_c$ is obtained from the $B-D$ mass splitting. }
\label{tab:central}
\end{table*}

\section{Conclusions}

In this talk I reviewed the current status of determining the magnitude of the CKM matrix elements $|V_{ub}|$ and $|V_{cb}|$ from inclusive semileptonic $B$ meson decays. 
For $B \to X_c \ell \bar \nu$, the operator product expansion has been calculated to order $1/m_b^3$, 
with a total of 6 parameters in addition to $|V_{cb}|$ appearing at that order. These 6 parameters
can be determined in a fit to precision measurements of inclusive decay spectra and 
one finds $|V_{cb}| = (42.1 \pm 0.6)\times 10^{-3}$. Also obtained in the fit is the value of the 
$b$-quark mass and the parameter $\lambda_1$, which are shown in Table~\ref{tab:central}. 

To measure $|V_{ub}|$ from the inclusive decay $B \to X_u \ell \bar\nu$ one has to deal with the large background from $b \to c$ transitions. Imposing kinematic cuts to suppress this background tends to destroy the convergence of the OPE. Several cuts have been presented which
allow to suppress this background experimentally, and in the future it should be possible
to determine the value of $|V_{ub}|$ with uncertainties well below the 10\% level. 


\section{Acknowledgements}
I would like to thank the organizers for organizing such a wonderful meeting in such a
wonderful location. This work was supported by the Department of Energy under
grant DE-FG03-92-ER-40701.

\end{document}